\documentclass[twocolumn,prb,amsmath,amssymb,showpacs]{revtex4}

\usepackage{graphicx}

\usepackage{dcolumn}

\usepackage{bm}

\newcommand{\tc}{$T_{\rm{C}}$}

\begin{document}
\title{Nanoengineered Curie Temperature in Laterally-patterned Ferromagnetic Semiconductor Heterostructures}
\author{K.\ F. \ Eid}
\author{B.\ L.\ Sheu}
\author{O.\ Maksimov}
\author{M.\ B.\ Stone}\altaffiliation{Condensed Matter Sciences Division, Oak Ridge National Laboratory, Oak Ridge TN 37831-6430.}
\author{P. \ Schiffer}
\author{N.\ Samarth}\email{nsamarth@psu.edu}
\affiliation{Dept. of Physics and Materials Research Institute, The Pennsylvania State University, University Park PA 16802}

\begin{abstract}
We demonstrate the manipulation of the Curie temperature of buried layers of the ferromagnetic semiconductor (Ga,Mn)As using nanolithography to enhance the effect of annealing. Patterning the GaAs-capped ferromagnetic layers into nanowires exposes free surfaces at the sidewalls of the patterned (Ga,Mn)As layers and thus allows the  removal of Mn interstitials using annealing. This leads to an enhanced Curie temperature and reduced resistivity compared to unpatterned samples. For a fixed annealing time, the enhancement of the Curie temperature is larger for narrower nanowires. 
\end{abstract}
\pacs{75.50 Pp, 75.75.+a, 81.16.-c}
\maketitle

Heterostructures derived from the ferromagnetic semiconductor (Ga,Mn)As are important for studying spin-dependent device concepts \cite{ohno_nature_1999,chiba_science_2003,tanaka_prl_2001,chun_prb_2002,chiba_prl_2004} relevant to semiconductor spintronics.\cite{wolf_science_2001,spinbook,samarth_SSP}  In contrast to metallic ferromagnets where crystalline defects rarely affect the magnetic ordering temperature, it is now well established that Mn interstitial defects -- double donors that compensate holes -- play a key role in limiting the hole-mediated Curie temperature (\tc) of (Ga,Mn)As to below 110 K in as-grown samples.\cite{yu_prb_2002} In thin (Ga,Mn)As epitaxial layers, post-growth annealing\cite{hayashi_apl_2001,potashnik_apl_2001} drives Mn interstitials to benign regions at the free surface where they are passivated, resulting in significant increases in both the hole density ($p$) and \tc~ (which can be as high as 160 K).\cite{edmonds_apl_2002,ku_apl_2003,edmonds_prl_2004} Such values of \tc~ should in principle allow the fabrication of proof-of-concept spintronic devices operating above liquid nitrogen temperatures. Unfortunately, the annealing-induced enhancement of \tc~ is almost completely suppressed in heterostructure devices containing buried (Ga,Mn)As layers.\cite{stone_apl_2003,chiba_apl_2003} In this Letter, we describe a nanoengineered solution to this problem: lithographic patterning of (Ga,Mn)As heterostructures into nanowires.  Such patterning opens up new pathways for defect diffusion to free surfaces at the sidewalls, and thus restores the annealing-enhanced \tc~ to buried (Ga,Mn)As layers.  These results suggest novel routes towards the flexible fabrication of (Ga,Mn)As-based spintronic devices with relatively high \tc.

Laterally-patterned wires are fabricated from GaAs/(Ga,Mn)As/GaAs heterostructures grown by molecular beam epitaxy on epiready semi-insulating GaAs (100) substrates. Crystal growth conditions are identical to those described in an earlier publication.\cite{stone_apl_2003} The samples consist of a 5 nm thick low temperature GaAs buffer layer, followed by a (Ga,Mn)As layer ($\sim 6\%$Mn, thickness of 15 nm or 50 nm) and finally a low temperature grown 10 nm GaAs capping layer. The data shown are for 50 nm thick magnetic layers, but the results were qualitatively the same for the 15 nm thick samples.  We pattern these samples into wires that are approximately 4 $\mu$m long and with two different widths (1 $\mu$m and 70 nm) using electron beam lithography followed by lift off and dry etching. 

For the patterning process, the samples are first spin coated with a bilayer of about 400-nm-thick electron-beam resist P(MMA-MAA) copolymer/PMMA  with molecular weight of 950. The desired patterns are defined on the sample using direct-write electron-beam lithography at electron energy of 100 keV. After development of the resist in MIBK:IPA 1:1 solution, a metallic layer of either 45 nm Al or Al/Au is deposited on the sample using thermal evaporation.  Using standard lift off techniques, we then obtain metal wires that serve as a hard mask for the subsequent chlorine-based dry etching process. After dry etching, for samples with an Al mask, the metal layer is dissolved in CD-26 photoresist developer, leaving just the semiconductor nanowire; for samples with an Al/Au mask, the metal is retained on portions of the sample. We note that in the latter case the undoped GaAs capping layer serves as an insulator that prevents the shunting of current through the metal during the measurement.\cite{note} Figures 1 (a) and (b) show plan view SEM images of patterned nanowires. Three identical rows of wires are patterned on each wafer: one set is measured as-grown and patterned, while the other two sets are annealed after patterning at $190^{\circ}$C for 5 hours.
 
We probe the ferromagnetic phase transition in single wires using four probe measurements of the temperature-dependent resistivity $\rho(T)$; it is well-established\cite{spinbook,samarth_SSP} that  $\rho(T)$ shows a well-defined peak close to \tc~ in (Ga,Mn)As, especially for samples with $T_{\rm{C}} < 100$ K. For higher \tc, the peak is less well-defined but still yields a reasonable estimate of the ordering temperature (typically an overestimate of $\sim 10$K). In addition, we measure the temperature-dependence of both the magnetization (using a commercial superconducting quantum interference device (SQUID) magnetometer) and the resistivity in macroscopic pieces of the parent wafers (using the van der Pauw method).

\begin{figure}[h]
\includegraphics[scale=0.7]{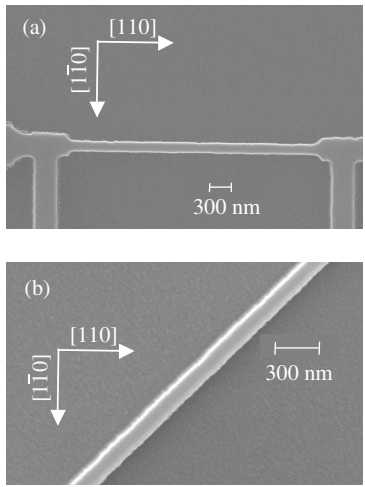}
\caption{SEM images of nanowires patterned along two different crystallographic directions: (a) [110] and (b) [010].}
\end{figure}

Figure 2 (a) shows the magnetization as a function of temperature for both as-grown and annealed ($\sim 5 \rm{mm}^2$) pieces of a GaAs/50 nm (Ga,Mn)As/10 nm GaAs heterostructure. As found in earlier studies,\cite{stone_apl_2003} the cap layer completely suppresses any annealing-induced enhancement of \tc. This suppression of the annealing effect has been attributed to the formation of a p-n junction at each of the two GaAs/(Ga,Mn)As interfaces as some Mn interstitials (double donors) initially diffuse across the boundaries into GaAs.\cite{edmonds_prl_2004,stone_apl_2003} The resulting Coulomb barrier is expected to strongly inhibit the further diffusion of Mn interstitials from the bulk of the (Ga,Mn)As layer towards the interfaces. The temperature-dependence of the sample resistance is in complete agreement with the magnetization measurements. Figure 2(b) shows Van der Pauw  measurements of $\rho(T)$ for a macroscopic mesa ($1 \rm{mm} \times 2 \rm{mm}$); the peak of $\rho(T)$  does not change upon annealing. In addition, there is no significant reduction in the resistivity of the sample with annealing, indicating that the hole density has not changed.

\begin{figure}[h]
\includegraphics[scale=0.7]{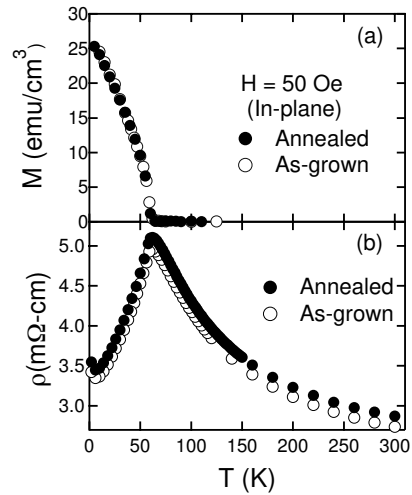}
\caption{(a) Magnetization versus temperature for macroscopic pieces (area of $\sim 5 \rm{mm}^2$) of as-grown and annealed GaAs/(Ga,Mn)As (50 nm)/ (10 nm) GaAs heterostructure.  The data are measured in a magnetic field of 50 Oe in-plane upon warming after field cooling at 10 kOe. The sample is annealed at $190^\circ$C in ultrahigh purity nitrogen gas (5N) for 5 hours. The data show that \tc~ is not affected by annealing.
(b) Resistivity versus temperature (measured in van der Pauw geometry) for a 1 mm $\times$ 2 mm mesa patterned from the same wafer as in (a).}
\end{figure}

Lateral patterning produces distinctly different behavior. Figure 3(a) shows the temperature-dependent resistivity for a $1 \mu$m wide wire patterned along the $[110]$ direction. The data are shown for both an as-grown wire and one annealed at $190^{\circ} \rm{C}$ for five hours. The data indicate that annealing results in a slight increase in \tc, accompanied by a slight increase in the resistivity of the sample. These observations suggest that annealing induces modest diffusion of Mn interstitials to the free surfaces at the sidewalls of the wire. The diffusion coefficient (estimated to be $D \sim 100 \rm{nm}^2 /\rm{hr}$ at $\sim 190^{\circ} \rm{C}$\cite{edmonds_prl_2004}) is simply not large enough to allow significant removal of Mn interstitials from the bulk of the $1 \mu{\rm{m}}$ wide wire within the five hours annealing time. There is however some alteration of the defect states, as indicated by the small increase in the high temperature resistivity; we do not have an explanation for this observation but note that a similar effect has been seen in previous studies of long anneals, albeit at higher annealing temperatures.\cite{potashnik_apl_2001} Figure 3 (b) shows the effect of annealing for a 70 nm wide nanowire (also patterned along $[1 1 0]$). Here, annealing produces a striking increase in \tc~ of almost 50 K, accompanied by a correspondingly significant decrease in the resistivity of the wire. Both these observations strongly suggest the successful removal of the Mn interstitials from the bulk of the (Ga,Mn)As nanowire. Lateral diffusion of the Mn interstitials provides the most likely explanation for these observations. We note that the time and length scales for diffusion observed in our experiments ($\sim 35$ nm in 5 hours) are consistent with earlier low-temperature annealing studies of (Ga,Mn)As epilayers.\cite{edmonds_prl_2004}

It is important to examine whether defect-diffusion may depend on the crystalline direction and/or the nature of the free surface where the interstitials eventually reside.   We hence pattern four nanowires (each with 70 nm width), along the principal cubic directions ( [110],[100],$[1 \overline{1} 0]$ and [010]). All four nanowires are annealed under identical conditions ($190 ^{\circ} \rm{C}$ for 5 hours). The results are shown in Fig. 3(c) and indicate that Ð at least for the processing and annealing protocol followed in this study Ð defect diffusion does not vary significantly with crystalline direction.

\begin{figure}[h]
\includegraphics[scale = 0.6]{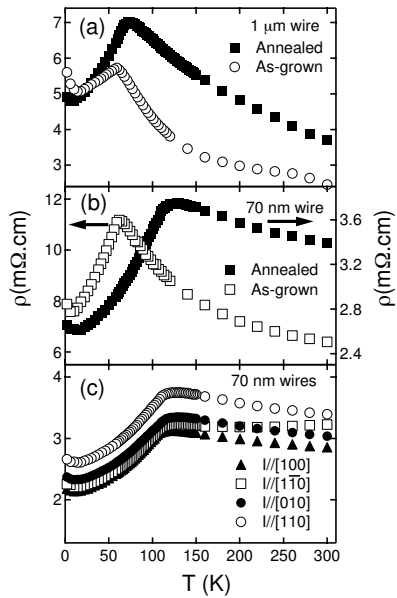}
\caption{Temperature dependence of the (four-probe) resistivity for as-grown and annealed single wires of (a) 1 micron width and (b) 70 nm width. Both sets of wires are patterned along the $[110]$ direction. Plot (c) shows the same measurements for four individual 70 nm wires patterned along different crystalline orientations. All wires are patterned from the same wafer used in Fig. 2, and the annealing conditions are identical to those used in Fig. 2.}
\end{figure}

In summary, we have shown that nanolithography allows the engineering of defect diffusion pathways, hence providing a means of tailoring impurity-controlled magnetism in ferromagnetic semiconductor heterostructures. Areas with different \tc~ and resistivity can be made on the same wafer simply by having different feature sizes. Smaller features have a higher Curie temperature and lower resistivity. Although we do not observe any obvious crystalline anisotropy in the diffusion constant of the Mn interstitials, such directional dependence may still exist and may be revealed by in situ resistance monitoring while annealing the nanowires. Our findings augur well for prospects of designing (Ga,Mn)As-based spintronic device heterostructures such as magnetic tunnel junctions with \tc~ well above 77 K. In addition, systematic studies of annealing of nanowires of differing width may provide new insights into ongoing controversies about the microscopic details of the diffusion and passivation of Mn interstitials in (Ga,Mn)As.\cite{adell_comment}
  
This research has been supported by grant numbers ONR N0014-05-1-0107, DARPA/ONR N00014-99-1093, -00-1-0951, University of California-Santa Barbara subcontract KK4131, and NSF DMR-0305238 and -0401486. This work was performed in part at the Penn State Nanofabrication Facility, a member of the NSF National Nanofabrication Users Network.


\begin{thebibliography}{18}
\bibitem{note} This is confirmed by comparing the resistivity of the wires with van der Pauw resistivity measured on samples that had no metal.
\end{thebibliography}

\end{document}